\documentclass[prd,floats,superscriptaddress,nofootinbib,preprintnumbers]{revtex4}%

\usepackage{lipsum, babel}

\usepackage{amssymb}
\usepackage{amsmath}

\usepackage{sans}

\makeatletter\AtBeginDocument{\let\@elt\relax}\makeatother

\usepackage{graphicx}
\usepackage{graphics}
\usepackage{dcolumn}
\usepackage{color}
\usepackage{rotate}
\usepackage{fancyhdr} 
\usepackage{hyperref}
\usepackage{indentfirst}
\usepackage{booktabs} 
\usepackage{fontawesome} 
\usepackage{physics} 
\usepackage{siunitx} 
\DeclareSIUnit \parsec {pc}
\DeclareSIUnit \year {yr}
\usepackage{umoline}

\def\be{\begin{eqnarray}}
\def\ee{\end{eqnarray}}
\def\ba{\begin{eqnarray}}
\def\ea{\end{eqnarray}}
\def\no{\nonumber}

\newcommand{\expect}[1]{\left\langle #1 \right\rangle}

\definecolor{darkred}{rgb}{.743,0,0}

\def\vphi{\varphi}
\def\d{\partial}
\def\x{{\bf x}}

\newcommand{\comment}[1]{}

\begin{document}
\title{A single-bubble source for gravitational waves in a cosmological phase transition}

\author{Kfir Blum}\email{kfir.blum@weizmann.ac.il}
\affiliation{Weizmann Institute of Science, Rehovot, Israel} 
\author{Mehrdad Mirbabayi}\email{mirbabayi@ictp.it}
\affiliation{International Centre for Theoretical Physics, Trieste, Italy} 

\begin{abstract}
We show that quantum fluctuations of an expanding phase transition bubble give rise to gravitational wave (GW) emission, even when considering a single bubble, without bubble collisions or plasma effects. The ratio of GW energy to the total bubble energy reservoir increases with time as $\propto t$. If the bubble expands for long enough before percolation destroys it, back-reaction due to the GW emission becomes important after $t_{\rm br}\sim (16\pi^5) m_{\rm pl}^2R_0^3$, where $R_0$ is the bubble nucleation radius and $m_{\rm pl}$ is the reduced Planck mass. As seen by experiments today, the GW energy spectrum would appear blue. 
However, simple estimates suggest that the signal falls short of detection by even ambitious future experiments.
\end{abstract}

\maketitle


\section{Introduction}\label{sec:int}
First-order cosmological phase transitions (PTs) are thought to proceed via the nucleation of bubbles of the true vacuum state~\cite{Coleman:1977py,Callan:1977pt,Coleman:1980aw}, and are considered a source of stochastic gravitational waves (GWs) with possible observational implications (see, e.g.~\cite{10.1093/mnras/218.4.629,Kamionkowski:1993fg}). Existing calculations focus on the collision of bubbles and plasma turbulence (for a review, see, e.g.~\cite{Weir:2017wfa}). This is because semiclassical estimates suggest that the dominant bubble configuration is spherical~\cite{Coleman:1977th,Blum:2016ipp}, and a spherical bubble cannot radiate GWs by itself. 
Here we point out that even if bubbles are spherical on average, quantum fluctuations lead to non-vanishing deformations. Thus GWs are produced also by a single bubble.  

To estimate the effect, we quantize the fluctuations on top of a thin-wall classical bubble, matching the fluctuation initial conditions to vacuum fluctuations of the state prior to nucleation. Solving for the time evolution of the fluctuation mode functions, we calculate the quantum root-mean-square (RMS) of the scalar field energy momentum tensor (EMT), and use this to compute the GW energy production. Our analysis applies under the assumption that the bubble size is much less than the Hubble distance.

We find that an expanding bubble converts a fraction $\alpha$ of its latent heat into GW energy, where $\alpha$ grows with time since nucleation as $\alpha\approx(t/t_{\rm br})$ with $t_{\rm br}\approx (16\pi^5)m_{\rm pl}^2 R_0^3$.
Here, $m_{\rm pl}=1/\sqrt{8\pi G}$ is the reduced Planck mass, and $R_0$ is the radius of the bubble at nucleation (typically comparable to the inverse of the energy scale characterizing the spontaneous symmetry breaking). 
The power spectral density (energy per logarithmic frequency bin) of the GWs as seen today peaks at the frequency $f_{\rm peak}\approx1/(R_0 (1+z_*))$, where $z_*$ is the redshift during the PT. 

Gravitational back-reaction may become significant around $t\sim t_{\rm br}$. Assuming the PT occurs in a radiation-dominated background with temperature $T_*$, the back-reaction time becomes shorter than a Hubble time for $(R_0m_{\rm pl})^3\lesssim 10^{-4}(m_{\rm pl}/T_*)^2$. In this regime, the classical treatment of bubble expansion and dynamics may be called into question.

Although we did not perform a survey of models, simple estimates suggest that the signal falls below the sensitivity of even the most ambitious currently planned GW experiments, like the Big-Bang Observer (BBO)~\cite{Moore:2014lga}. 

The theory of domain wall fluctuations was developed long ago in Ref.~\cite{Vilenkin}, though to our knowledge the resulting GW emission has not been worked out. There have been several works on the quantum state of the system after bubble nucleation~\cite{Sasaki:1993ha,Sasaki:1993gy,Tanaka:1994qa}, and on the spectrum of cosmological perturbations (scalars and tensors) as measured by the Friedmann–Lemaître–Robertson–Walker (FLRW) observers residing inside one expanding bubble~\cite{Yamamoto:1994te,Yamamoto:1996qq,Garriga}. However, the map between those results and our current computation seems quite complex, since from the global perspective the unperturbed bubble itself breaks homogeneity and isotropy of the Minkowski slices, so scalar fluctuations of the bubble wall source GWs at linear order; in contrast, on the open FLRW slices inside the bubble, scalar and tensor perturbations are decoupled at linear order. 

\section{Bubble-wall fluctuations}
Semi-classically, the decay of a false vacuum in 4d is dominated by an O(4)-symmetric Euclidean saddle~\cite{Coleman:1977py}. In real time, it describes the nucleation of a spherically symmetric bubble with a profile that is solely a function of $R$ in the following parametrization of Minkowski spacetime,
\be\label{eq:coord} ds^2&=&dR^2+R^2\left(-d\tau^2+\cosh^2\tau d\Omega^2\right).\ee
The $R,\tau$ coordinates cover the region of spacetime located outside of the lightcone centered on the bubble. In the thin-wall regime~\cite{Coleman:1977py}, this is where interesting dynamics happens, with an expanding region of true vacuum separated from the false vacuum by a wall at $R=R_0$. In terms of the wall tension $\sigma$ and the difference $\epsilon$ between the vacuum energy density in the true vs. false vacua,
\be\label{eq:R0} R_0&=&\frac{3\sigma}{\epsilon}.\ee

Of course, any realization of this process exhibits quantum fluctuations around the symmetric bubble. Our goal in this paper is to calculate GW production by these fluctuations. In general, this is a challenging task; we will simplify the problem by focusing on the GW production in the regime $t\gg R_0$, after the bubble expands beyond its radius at nucleation. The idea is that the long-lived fluctuations of the bubble configuration in the thin wall regime are the transverse excitation modes, which can be described by a 3d effective field theory (EFT).

The dynamics of transverse excitations of a brane is governed by the Nambu-Goto (or Dirac-Born-Infeld in the case of codimension 1) effective action at leading order in the derivative expansion. In addition, the domain wall between the false and true vacua experiences pressure. The action in Euclidean time is therefore the 3d area minus the 4d volume:
\be\label{eq:Sdef} S&=&A_3\sigma-V_4\epsilon.\ee
For the O(4)-symmetric bubble of radius $R$, this action is $S=2\pi^2\left(\sigma R^3-\frac{\epsilon}{4}R^4\right)$, whose stationary point is given by Eq.~(\ref{eq:R0}).

We write the action in static gauge, in which the wall world-volume is parameterized by $y^\alpha=(\tau,\theta,\varphi)$, the same coordinates that appear in Eq.~(\ref{eq:coord}), and parameterize the fluctuations by
\be R&=&R_0e^\zeta.\ee
The action Eq.~(\ref{eq:Sdef}) becomes
\be\label{eq:Sfluc} S&=&\sigma R_0^3\int d^3y\sqrt{g}\left(e^{3\zeta}\sqrt{1+g^{\alpha\beta}\partial_\alpha\zeta\partial_\beta\zeta}-\frac{3}{4}e^{4\zeta}\right),\ee
where $g_{\alpha\beta}$ is the metric of a unit 3-sphere. It analytically continues to the metric of dS$_3$, which is multiplying $R^2$ in Eq.~(\ref{eq:coord}). We are interested in the regime $\sigma R_0^3\gg1$, in which $\zeta$ is weakly coupled. The quadratic action for $\zeta$ reads
\be\label{eq:S(2)} S^{(2)}&=&\frac{1}{2}\sigma R_0^3\int d^3y\sqrt{g}\left(g^{\alpha\beta}\partial_\alpha\zeta\partial_\beta\zeta-3\zeta^2\right).\ee
In Lorentz signature, $\zeta$ has tachyonic mass. This will lead to a growing mode solution, which we will analyze in detail.

The EMT of the bubble deviates from spherical symmetry, already at linear order in $\zeta$. Hence, the power spectrum of GWs can be determined from the power spectrum of $\zeta$. This follows from a standard inflationary computation. We expand $\zeta$ in spherical harmonics on the 2-sphere,
\be\zeta(\tau,\theta,\varphi)&=&\sum_{l,m}\zeta_{lm}(\tau)Y_{lm}(\theta,\varphi),\ee
and quantize each $\zeta_{lm}$
\be\zeta_{lm}(\tau)&=&f_l(\tau)a_{lm}+f^*_l(\tau)a^\dag_{lm},\;\;\;\;\;\;[a_{lm},a^\dag_{l'm'}]=\delta_{ll'}\delta_{mm'}.\ee
The mode function $f_l(\tau)$ is a solution of
\be\partial_\tau\left(\cosh^2\tau\partial_\tau f_l(\tau)\right)+\left(l(l+1)-3\cosh^2\tau\right)f_l(\tau)&=&0.\ee
If we analytically continue $\tau\to-i\tau$, the bubble wall becomes a 3-sphere. We choose the state of $\zeta_{lm}$ to be the Hartle-Hawking vacuum, given by $a_{lm}|0\rangle=0$ if $f_l(\tau)$ is chosen to be the solution that is regular on this sphere~\cite{Hartle:1983ai}. 
This gives 
\be f_l(\tau)&=&\frac{N_l}{4}\frac{2Q_{l,2}\left(\tanh\tau\right)+i\pi P_{l,2}\left(\tanh\tau\right)}{\cosh\tau},\ee
where $N_l$ is a normalization constant and $P_{l,m}$ and $Q_{l,m}$ are the associated Legendre functions of first and second kind. 
The normalization is fixed by the canonical commutation relation between $\zeta_{lm}$ and its conjugate momentum $\pi_{lm}=\frac{\d\mathcal{L}}{\partial(\d_\tau\zeta_{lm})}=\sigma R_0^3\cosh^2\tau\d_\tau\zeta_{lm}$. Requiring $[\zeta_{lm},\pi_{l'm'}]=\sigma R_0^3\cosh^2\tau\left(f_l\d_\tau f^*_{l'}-\d_\tau f_{l'} f^*_l\right)\delta_{ll'}\delta_{mm'}=i\delta_{ll'}\delta_{mm'}$ at $\tau\to0$, we have
\be\label{eq:Nl}N_l&=&\frac{2}{\sqrt{\pi\sigma R_0^3}}\sqrt{\frac{\Gamma\left(l-1\right)}{\Gamma\left(l+3\right)}}.
\ee
%


Because of the tachyonic mass, the mode function grows exponentially at late times,
\be\label{eq:flate} f_l(\tau)\to\frac{N_l}{2}e^\tau\left(1+\left(l(l+1)-1\right)e^{-2\tau}+...\right).\ee
At large $l$ this goes as
\be f_l(\tau)\approx\frac{e^{\tau}}{\sqrt{\pi\sigma R_0^3}l^2},\;\;\;\;{\rm for}~~\tau\gg\log l\gg1,\ee
which is dictated by dS$_3$ dilation isometry. Note however that the expansion in $\zeta$ breaks down when
\be\label{eq:tauNL} \tau&\gg&\tau_{\rm NL}=\ln\sqrt{\sigma R_0^3},\ee
corresponding to Minkowski time 
\be\label{eq:tNL} t_{\rm NL}&=&\frac{R_0}{2}\sqrt{\sigma R_0^3}.\ee
It is important to understand the physical meaning of the growth of $f_{l}$, and we consider this point next. 

\subsection{Nonlinear solution}\label{sec:nl}
The exponential growth of $\zeta$ at superhorizon scales has a simple explanation.\footnote{Courtesy of M. Zaldarriaga.} It corresponds to elements of the wall approaching the speed of light. In this limit, a displacement from the unperturbed trajectory remains frozen as viewed in Minkowski coordinates, related to the hyperbolic coordinates via
\be\label{eq:MinkHyp} t&=&R\sinh\tau,\;\;\;{\bf x}=R\cosh\tau\,{\bf \hat n}.\ee
For a small perturbation $\zeta\ll1$, let us evaluate the equal-$t$ displacement $\Delta r(\tau)=|\x(\tau)|-|\x_0(\tau_0(\tau))|$, where $\tau_0$ is determined by
\be R\sinh\tau&=&R_0\sinh\tau_0=t.\ee
For $\tau\gg1$, we find
\be\label{eq:Dx}\Delta r&=&e^{-\tau}\left(R-\frac{R_0^2}{R}\right)\left(1+\mathcal{O}\left(e^{-2\tau}\right)\right).\ee
When $\zeta\ll1$ this simplifies to
\be\Delta r&\approx&2R_0\zeta e^{-\tau}\left(1+\mathcal{O}\left(e^{-2\tau}\right)\right).\ee
Thus, the exponential growth of the superhorizon fluctuations of $\zeta$ in terms of $\tau$, at least at linear order, has the simple interpretation that when $t\gg R_0$, different surface elements of the wall follow approximately null radial trajectories with small constant offsets $r(t)=t+\Delta r_0+\mathcal{O}\left(R_0^2/t^2\right)$. 

The above interpretation suggests that the exponential growth of $\zeta$ as viewed in hyperbolic coordinates is not a sign of any instability, but rather corresponds to $\Delta r$ approaching a constant at $t>t_{\rm NL}$. This can be verified in the regime $\zeta\gg1$, by using the nonlinear equation of motion
\be-\frac{1}{\cosh^2\tau}\partial_\tau\left(\frac{\cosh^2\tau R^3\partial_\tau R}{\sqrt{R^2+\left(\partial R\right)^2}}\right)+\frac{1}{\cosh^2\tau}\nabla_a\left(\frac{R^3\nabla^aR}{\sqrt{R^2+\left(\partial R\right)^2}}\right)-3\left(R^2\sqrt{R^2+\left(\partial R\right)^2}-\frac{R^4}{R_0}\right)&=&0,\ee
where $\nabla_a$ is the covariant derivative on S$^2$. The growing solution of this equation is of the form
\be R&=&c_1e^\tau+\frac{R_0^2}{c_1}e^{-\tau}+\mathcal{O}\left(e^{-3\tau}\right).\ee
In particular, the angular dependence only affects the $\mathcal{O}\left(e^{-3\tau}\right)$ terms. Substituting in Eq.~(\ref{eq:Dx}), one finds $\Delta r=\Delta r_0+\mathcal{O}\left(e^{-2\tau}\right)$. 

In App.~\ref{app:chi} we use Minkowski coordinates to parameterize the wall. Although these coordinates are less adapted to the symmetry of the problem, the analysis manifests that fluctuations freeze rather than grow at late times. 

\subsection{Energy momentum tensor}
In the thin-wall approximation, and up to a cosmological constant, the EMT is the sum of the wall tension and a bulk contribution from the difference between the vacuum energies. For the purpose of calculating GW emission, we can express the EMT in Minkowski coordinates $x^\mu=(t,{\bf x})$,
\be T^{\mu\nu}&=&\sigma h^{\alpha\beta}\partial_\alpha x^\mu\partial_\beta x^\nu\delta\left(\xi\right)-\frac{3\sigma}{R_0}\eta^{\mu\nu}\theta\left(-\xi\right),\ee
where $h^{\alpha\beta}$ is the inverse of the induced metric on the wall and $\xi$ is a coordinate that is normal to the wall and vanishes at its location. Using $\tanh\tau=t/r$, we have
\be\label{eq:taud}\partial_t\tau&=&\frac{\cosh^2\tau}{r},\;\;\;\partial_i\tau=-\frac{\cosh^2\tau}{r^3}tx^i.\ee
The wall is localized at $\sqrt{{\bf x}^2-t^2}=R_0e^\zeta$. 
Using this and Eq.~(\ref{eq:taud}), up to $\mathcal{O}\left(\zeta\right)$ we have
\be \label{eq:dxi}d \xi&=&\left(-\frac{t}{R_0}(1-\zeta)-\partial_t\zeta,\frac{x^i}{R_0}(1-\zeta)-\partial_i\zeta\right).\ee
From Eq.~(\ref{eq:MinkHyp}), the induced metric is
\be h_{\alpha\beta}&=&R_0^2e^{2\zeta}\left(g_{\alpha\beta}+\partial_\alpha\zeta\partial_\beta\zeta\right),\ee
where $g_{\alpha\beta}$ is the metric of dS$_3$ with unit radius of curvature. 
%
%
Up to $\mathcal{O}\left(\zeta\right)$, we find
\be \label{eq:T00}T^{00}&=&\sigma\left(\cosh^2\tau+\sinh2\tau\partial_\tau\zeta\right)\delta\left(\xi\right)-\frac{3\sigma}{R_0}\theta\left(-\xi\right),\\
\label{eq:T0i}T^{0i}&=&\sigma\left[\left(\sinh\tau\cosh\tau+\cosh2\tau\partial_\tau\zeta\right)\hat x^i-\tanh\tau\left(\partial_\theta\zeta\hat\theta^i+\frac{\partial_\varphi\zeta}{\sin\theta}\hat\varphi^i\right)\right]\delta\left(\xi\right),\\
\label{eq:Tij}T^{ij}&=&\sigma\left[-\delta^{ij}+\left(\cosh^2\tau+\sinh2\tau\partial_\tau\zeta\right)\hat x^i\hat x^j-\partial_\theta\zeta\left(\hat x^i\hat\theta^j+\hat x^j\hat\theta^i\right)-\frac{\partial_\varphi\zeta}{\sin\theta}\left(\hat x^i\hat\varphi^j+\hat x^j\hat\varphi^i\right)\right]\delta\left(\xi\right)+\frac{3\sigma}{R_0}\delta^{ij}\theta\left(-\xi\right).\no\\&&
\ee
In practice, we will need to integrate $T^{\mu\nu}$ on a constant-$t$ slice. Hence, $\tau={\rm arctanh}\left(t/r\right)$ and $\theta(-\xi)$ and $\delta(\xi)$ will introduce additional $\zeta$ dependence. 

One can check that
\be\int d^3x\partial_\mu T^{\mu\nu}&=&\int d\Omega\left({\rm EOM\;of\;\zeta}\right).\ee
%

\section{Gravitational waves}\label{sec:GW}
At linear order, the transverse-traceless gravitational waves satisfy~\cite{Maggiore:2007ulw}
\be\ddot h_\pm(t,{\bf k})+k^2h_\pm(t,{\bf k})&=&16\pi G\varepsilon_\mp^{ij}\left(\hat k\right)T_{ij}\left(t,{\bf k}\right),\ee
where the polarization tensor of $\pm2$ helicity waves for ${\bf k}\propto\hat z$ is given by
\be\varepsilon_\pm^{ij}&=&\frac{1}{2}\left(\hat x\pm i\hat y\right)^i\left(\hat x\pm i\hat y\right)^j.\ee
The solutions are 
\be\label{eq:hpm} h_\pm\left(t,{\bf k}\right)&=&16\pi G\int_0^tdt'\frac{\sin k(t-t')}{k}T_\pm\left(t',{\bf k}\right),\ee
with the source
\be \label{eq:Tpm}T_\pm\left(t,{\bf k}\right)&=&\int d^3xe^{i{\bf kx}}\varepsilon_\mp^{ij}T_{ij}\left(t,{\bf x}\right).\ee
In Eq.~(\ref{eq:hpm}) we take the lower limit of the integration to be $t'=0$, which is the moment of the materialization of the bubble. 

Only the surface tension part of $T_{ij}$ in Eq.~(\ref{eq:Tij}) gives a nonzero contribution when contracted with $\varepsilon^{ij}$. Focusing on the $(+)$ polarization and integrating over the radial direction using the $\delta(\xi)$, we find up to $\mathcal{O}\left(\zeta\right)$
\be\label{eq:Tpm2}T_+\left(t,{\bf k}\right)&=&-\frac{\sigma R_0r_w}{2}\int d\theta\sin^3\theta\,\int d\varphi\,e^{ikr_w\cos\theta - 2i\varphi}\left(-\frac{r_wt}{R_0^2}\d_\tau\zeta
+\left(4+\frac{t^2}{R_0^2}+ikr_w\cos\theta\right)\zeta
\right).\ee
The unperturbed spherical bubble term drops out in the angular integration. We integrated by parts in $\theta$ and $\varphi$ to eliminate derivatives, and used 
\be r\partial_r\zeta=-\frac{rt}{R^2}\partial_\tau\zeta.\ee
In \eqref{eq:Tpm2}, $r_w=\sqrt{R_0^2+t^2}$ defines the unperturbed bubble wall position. 

Expanding $\zeta$ in $Y_{lm}$, and performing the $\varphi$ integral, we have
\be \!\!\!\!\!T_+\left(t,{\bf k}\right)&=&-\frac{\sigma R_0r_w}{2}\sum_{l}\sqrt{\frac{2l+1}{4\pi}\frac{\Gamma\left(l-1\right)}{\Gamma\left(l+3\right)}}\int_{-1}^1 dc(1-c^2)\,e^{ikr_wc}P_{l,2}\left(c\right)\left(\left(4+\frac{t^2}{R_0^2}+ikr_wc\right)\zeta_{l,2}-\frac{r_wt}{R_0^2}\d_\tau\zeta_{l,2}
\right).\ee
This expression is difficult to handle in general, so in what follows we focus on the large-$t$ region. 
\subsection{Large $t$ expansion}
We now consider the limit $t\gg R_0$, with fixed $kt$. In this limit we can use the fact that $f_l(\tau)$ is real to $\mathcal{O}\left(R_0^3/t^3\right)$, to write 
\be\zeta_{lm}&=&(a_{lm}+a^\dag_{lm})f_l+\mathcal{O}\left(R_0^3/t^3\right)\ee
and expand $f_l$ and its $\tau$ derivative $\d_\tau f_l$ as
\be\label{eq:flateA} f_l=\frac{N_l\,t}{R_0}\left(1+l(l+1)\frac{R_0^2}{4t^2}+...\right)\,,\;\;\;\;\;\frac{df_l}{d\tau}= \frac{N_l\,t}{R_0}\left(1-l(l+1)\frac{R_0^2}{4t^2}+...\right).\ee
With these approximations we find
\be\label{eq:Tp}
T_{+,l}\left(t,{\bf k}\right)&=& c_T(l) \left(a_{l,2}+a^\dag_{l,2}\right) \frac{I_l(kt)}{k^2},\no\\
c_T(l) & \equiv & -\sqrt{\frac{(2l+1) \sigma}{R_0^3}}\frac{\Gamma\left(l-1\right)}{4\pi\Gamma\left(l+3\right)},\\
I_l(w)& \equiv & w^2 \int_{-1}^1 dc\,(1-c^2)\,e^{ic w}P_{l,2}\left(c\right)\left(2ic w+l(l+1)+6\right).\no
\ee

An important observation to make is that the late time growth of the integrand in Eq.~(\ref{eq:Tp}), at fixed $kt$, is $\sim t^2$, because the terms of order $t^4$ cancel. This naive growth and absence thereof can be understood as follows. The total energy in the bubble wall is
\be E_{\rm wall}&\sim&\epsilon t^3\sim\frac{\sigma}{R_0}t^3.\ee
The GW source term Eq.~(\ref{eq:Tpm}) is also a volume integral of the energy-momentum tensor, but because of the spherical symmetry of the bubble there is no GW production at zeroth order in $\zeta$. So, naively, one could expect $T_+(t,{\bf k})\sim\frac{\sigma}{R_0}t^3\zeta\sim t^4$. On the other hand, we saw that the leading growth of $\zeta$ accounts for the radial motion of the wall elements at the speed of light, but with fixed displacement $\Delta r_0$. Therefore, up to the inverse radius of curvature of the wall, which is $\mathcal{O}\left(1/t\right)$, and up to $k$ times this displacement $\Delta r_0\sim R_0^2\zeta/t$, we can move the surface elements back to the original, spherically-symmetric positions and get a vanishing result.

Proceeding with the calculation, we can then perform the time integral of Eq.~(\ref{eq:hpm}), \footnote{The integration in Eq.~(\ref{eq:hpmA}) includes a segment that lies outside the domain of validity of the large-$t$ approximation. One can verify, however, that cutting-off the integral below $w\approx kR_0$ does not affect the final results substantially (but leads to more complicated expressions that we prefer to avoid).}
\be\label{eq:hpmA} h_{+,l}\left(t,{\bf k}\right)&=& 16\pi G c_T(l)\frac{H_l(kt)}{k^4} \left(a_{l,2}+a^\dag_{l,2}\right),\\
H_l(x) &\equiv& \int_0^{x}dw \sin(x-w) I_l(w).\no\ee
The expectation value of energy in GWs (with a factor of 2 to include the $(-)$ helicity) follows to be
\be\label{eq:El}
\left\langle E_{{\rm GW},l}(t)\right\rangle &=& \frac{1}{32\pi G} 
\int \frac{d^3k}{(2\pi)^3} \expect{|\dot h_{+,l}\left(t,{\bf k}\right)|^2 +k^2 |h_{+,l}\left(t,{\bf k}\right)|^2}\no\\[10 pt]
& = & \frac{4G}{\pi} c_T^2(l) t^3 \int_0^\infty \frac{ds}{s^4}[|H_l(s)|^2+|\partial_sH_l(s)|^2].
\ee
For each $l<t/R_0$, this integral is dominated by $s = kt\sim l$. In fact, although we did not find a general expression, $l$-by-$l$ we can evaluate the integral analytically. For instance, setting $l=2$, we find 
\be \!\!\!\!\!\!\!\!\!\!T_{+,l=2}\left(t,{\bf k}\right)&=&-\left(a_{2,-2}+a^\dag_{2,-2}\right)\sqrt{\frac{\sigma}{R_0^3}}\frac{\sqrt{5}}{\pi k^2}\frac{w(w^2+3)\cos w - 3\sin w}{w^3}
\ee
and
\be
\!\!\!\!h_{+,l=2}\left(t,{\bf k}\right)& = &-\left(a_{2,-2}+a^\dag_{2,-2}\right)\frac{8\sqrt{5}G}{k^4}\sqrt{\frac{\sigma}{R_0^3}}\Big(3\cos w+\frac{(w^2-3)\sin w}{w}\Big),
\ee
leading to 
\be \left\langle E_{{\rm GW},l=2}(t)\right\rangle &=&\frac{22}{21\pi^2}\frac{G\sigma t^3}{R_0^3}. 
\ee
%
Similarly, every $l<t/R_0$ contributes to the GW energy $\propto t^3$. The large $t$ behavior of total $E_{\rm GW}$, therefore, depends on the large $l$ behavior of $E_{{\rm GW},l}$. This and other asymptotics will be analytically derived in appendix \ref{app:l}. 

\begin{figure}[ht!]
\centering
\includegraphics[width=0.6\hsize]{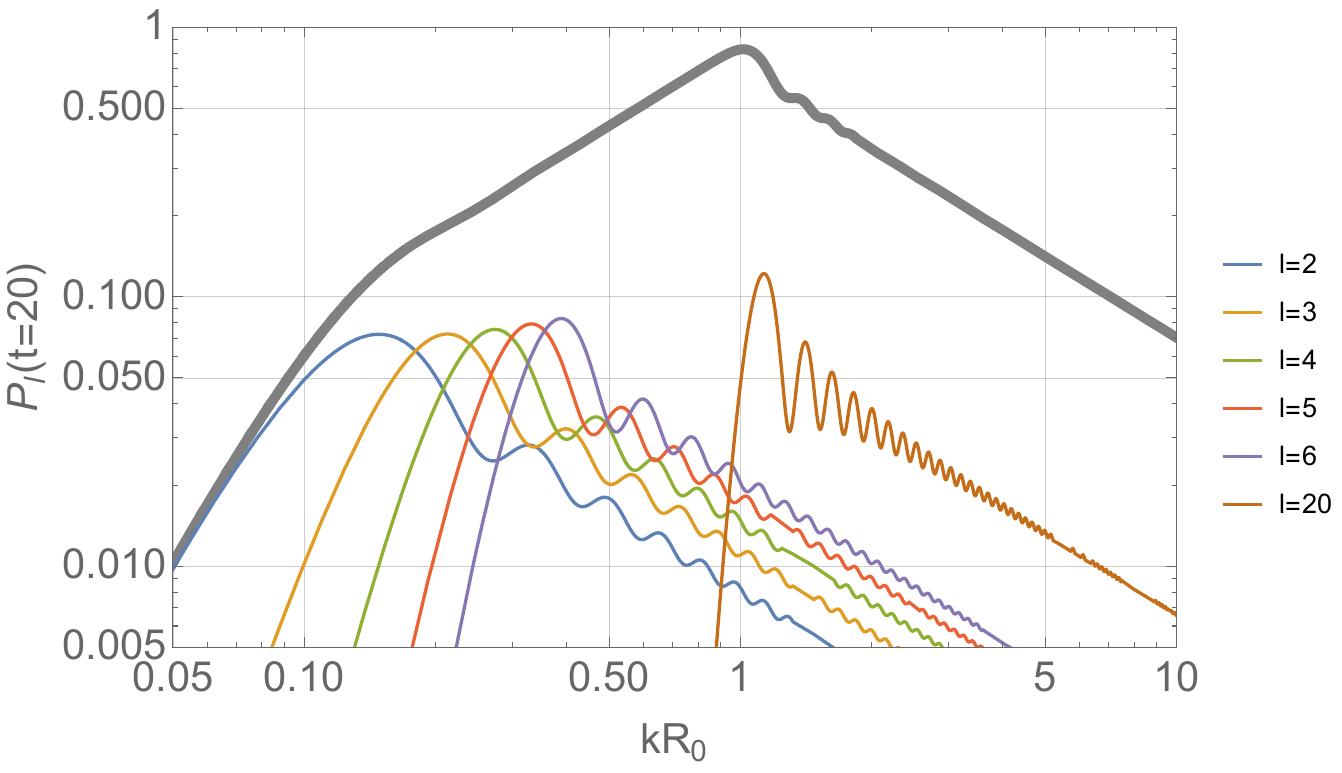}
\caption{Thin lines: $P_{l}$ for a few values of $l$, evaluated at $t=20R_0$. Thick line: sum over $l$ up to $l=20=t/R_0$. Note that the large-$t$ expansion breaks down for $l>t/R_0$.}\label{fig:Pls}
\end{figure}

We define $P_l$ by
\be \left\langle E_{{\rm GW},l}(t)\right\rangle&=&\frac{G\sigma t^3}{R_0^3}\int_0^\infty d\ln s\,P_{l}(s).
\ee
The computation can be trusted up to $l\approx t/R_0$, beyond which the large $t$ expansion breaks down. 
Results for a few values of $l$, evaluated at $t=20R_0$, are shown in Fig.~\ref{fig:Pls}. 
The peak of the power spectrum for each separate $l$ contribution occurs at $s\approx l$. For large $s\gg l$, $P_l(s) \approx B_l \frac{l}{s}$, with $B_l$ approaching $\frac{2}{\pi^3}\approx 0.065$ at large $l$. For small $s<l$ there is a steep cut-off $P_l(s)\propto s^{2l-1}$. The peak amplitude depends only weakly on $l$, and we find that all modes have $\left\langle E_{{\rm GW},l}(t)\right\rangle\approx A_l\frac{G\sigma t^3}{R_0^3}$, with the numerical coefficient $A_l$ converging slowly to $A_\infty = \frac{3}{4\pi^2} \approx 0.076$. For example, for $l=\{2,5,20,60\}$ we have $A_l = \left\{\frac{22}{21 \pi ^2},\frac{94}{117 \pi ^2},\frac{1264}{1677 \pi ^2},\frac{10984}{14637 \pi ^2}\right\}=\{0.1061,0.0814,0.0763,0.0760\}$.  

To find the total GW spectrum, using the above asymptotics we can estimate the sum over $l$ contributions up to $l\approx t/R_0$ by
\be\label{eq:Pk} P(k)&\approx &\sum_{l=2}^{t/R_0}P_l(s=kt)
\approx \frac{2}{\pi^3} 
\left\{\begin{array}{cc} k^3 t^3/8, \qquad &k<2/t,\\[10 pt]
kt/2,\qquad&2/t <k<1/R_0,\\ [10pt]
t/(2 k R_0^2),\qquad & 1/R_0<k\end{array}\right.\ee
%
With those approximations and assumptions, the total energy in GWs produced by the bubble is
\be\label{eq:EgwTotal} \left\langle E_{{\rm GW}}(t)\right\rangle&=&\frac{G\sigma t^3}{R_0^3}\int \frac{dk}{k}P(k)\approx \frac{2}{\pi^3} \frac{G\sigma t^4}{R_0^4}.\ee
%

\section{Discussion}\label{sec:disc}

It is useful to consider the ratio between the GW energy and the latent heat of the bubble. Using Eq.~(\ref{eq:EgwTotal}) we find
\be\alpha_{\rm GW}(t)&\equiv&\frac{E_{\rm GW}}{\frac{4\pi}{3}\epsilon t^3}\approx \frac{8\pi G}{16\pi^5R_0^2}\frac{t}{R_0},\ee
where we used $R_0=3\sigma/\epsilon$.
This result suggests that the GW energy overcomes the total energy budget of the bubble within a back-reaction time defined via $\alpha_{\rm GW}(t_{\rm br})=1$, and given by
\be t_{\rm br}&=&16\pi^5m_{\rm pl}^2R_0^3.\ee

We will consider the duration of the PT, $t_{\rm PT}$, as a free parameter. 
For our estimates below, it is useful to define $t_{\rm PT}$ as a fraction of the Hubble time,
\be t_{\rm PT}&=&f_{\rm PT}\,t_H,\;\;\;\;\;\;\;\;f_{\rm PT}<1.
\ee
If the PT takes place in a radiation-dominated background with temperature $T_*$ and $g_*$ effective relativistic degrees of freedom (dof), the Hubble time $t_H$ is
\be t_H^{-1}&=&H_*=\sqrt{\frac{\pi^2g_*}{90}}\frac{T_*^2}{m_{\rm pl}}\approx3\,\left(\frac{g_*}{100}\right)^{\frac{1}{2}}\frac{T_*^2}{m_{\rm pl}},\ee
so the ratio of the back-reaction time to the PT duration is
\be\label{eq:tbrtH}\frac{t_{\rm br}}{t_{\rm PT}}&\approx&\frac{1.6\times 10^4}{f_{\rm PT}}\,\left(\frac{g_*}{100}\right)^{\frac{1}{2}}\left(R_0T_*\right)^2\left(R_0m_{\rm pl}\right).\ee
Our field theory analysis can only be trusted for $R_0m_{\rm pl}\gg1$. However, we have no obvious a-priori expectation for the product $R_0T_*$. As long as $(R_0T_*)\gtrsim0.01f_{\rm PT}^{\frac{1}{2}}/\sqrt{R_0m_{\rm pl}}$, the back-reaction time is longer than the duration of the PT and back-reaction can probably be neglected for the purpose of estimating the GW energy, and in the usual analysis of bubble dynamics. Note that this would indeed be the case if $R_0\sim T_*$, which may be natural if thermal effects are important in triggering the phase transition.

The ``strength of the PT" is often parameterized by the ratio of the latent heat energy density to the background total energy density,
\be\label{eq:aPT}\alpha_{\rm PT}&\equiv&\frac{\epsilon}{3m_{\rm pl}^2H_*^2}.\ee
With this, the fraction of GW energy density to the total energy density at the time of the PT is
\be\Omega_{\rm GW*}&=&\alpha_{\rm GW}(t_{\rm PT})\,\alpha_{\rm PT}=\alpha_{\rm PT}\frac{t_{\rm PT}}{t_{\rm br}}.\ee
The fraction of GW energy density observable today is
\be\Omega_{\rm GW}&=&\frac{1}{(1+z_*)^4}\frac{H_*^2}{H_0^2}\Omega_{\rm GW*}
\ee
where $H_0=100h$~km/s/Mpc, $1+z_*=(g_*/g_{s0})^{\frac{1}{3}}(T_*/T_0)$, $g_{s0}\approx3.9$ is the effective number of relativistic dof contributing to entropy today at $T_0\approx2.7$~K, and we assumed $g_*\approx g_{s*}$.
Altogether, we find
\be\label{eq:Ogwh2}\Omega_{\rm GW}h^2
&\approx&4\times10^{-12}\,\alpha_{\rm PT}\,f_{\rm PT}\,\left(\frac{100}{g_*}\right)^{\frac{5}{6}}\left(\frac{1}{R_0T_*}\right)^3\left(\frac{T_*}{10^{16}{\rm GeV}}\right)\\
&\approx&1.5\times10^{-5}\left(\frac{100}{g_*}\right)^{\frac{1}{3}}\alpha_{\rm PT}\frac{t_{\rm PT}}{t_{\rm br}}.\no
\ee
For comparison, cosmic microwave background constraints on excess relativistic species imply $\Omega_{\rm GW}h^2\lesssim10^{-6}$~\cite{Smith:2006nka}. From the second line of Eq.~(\ref{eq:Ogwh2}), our model cannot probe this constraint without significant back-reaction, in a regime where our treatment is not under good control. \footnote{The constraint on excess relativistic species~\cite{Smith:2006nka} still allows a contribution to the energy density that is in the same order of magnitude as that of the photons. Therefore, the fact that GWs from cosmological PTs cannot saturate this bound without gravitational back-reaction, should also apply to bubble collisions.} 

Considering the frequency spectrum, GWs emitted at frequency $f_*$ during the phase transition are observed today at redshifted frequency $f$ given by 
$f=f_*/(1+z_*)$. 
The peak frequency at emission is $f_{*,\rm peak}\approx1/R_0$, so the observed peak frequency today is 
\be f_{\rm peak}&\approx&120\left(\frac{100}{g_*}\right)^{\frac{1}{3}}\frac{1}{R_0T_*}~{\rm GHz}.\ee
For $f<f_{\rm peak}$, the second spectral feature of the signal is the transition from $P(k)\propto k^3$ to $P(k)\propto k$, which happens at a frequency $f_{*,\rm break}\approx\frac{2}{t_{\rm PT}}=\frac{2H_*R_0}{f_{\rm PT}}f_{*,\rm peak}$, redshifted today into 
\be\label{eq:fbreak} f_{\rm break}&\approx&\frac{3}{f_{\rm PT}}\left(\frac{100}{g_*}\right)^{\frac{1}{6}}\frac{T_*}{10^{16}~{\rm GeV}}~{\rm GHz}.\ee
Current and future planned GW detectors are geared towards much lower frequencies than the illustrative values we use here for $f_{\rm peak}$~\cite{Moore:2014lga}. The spectrum would therefore appear blue. 
For $f_{\rm break}<f<f_{\rm peak}$ we can write:
\be \frac{d\Omega_{\rm GW}h^2}{d\ln f}&\approx&\frac{\Omega_{\rm GW}h^2}{2}\frac{f}{f_{\rm peak}}\\
&\approx&5\times 10^{-19}\,\left(\frac{100}{g_*}\right)^{\frac{1}{4}}\alpha_{\rm PT}\left(f_{\rm PT}\,\frac{t_{\rm PT}}{t_{\rm br}}\right)^{\frac{1}{2}}\left(\frac{1}{R_0\,m_{\rm pl}}\right)^{\frac{1}{2}}\frac{f}{\rm Hz}.\no
\ee
We arranged the last line such that all of the parameter combinations are smaller than one.
For comparison, the optimal sensitivity of the planned BBO experiment may reach $\Omega_{\rm GW}h^2\approx10^{-14}$ at $f\approx0.1$~Hz~\cite{Moore:2014lga}.

\section{Summary}
We analyzed the gravitational waves (GWs) produced by an expanding phase transition bubble in the thin-wall regime. Quantum fluctuations around the classical bubble configuration lead to nontrivial angular moments for the bubble's energy momentum tensor, sourcing GWs from a single bubble even before the onset of bubble collisions.

The total GW energy produced by the bubble grows with time as $E_{\rm GW}\propto t^4$. As a result, the ratio between the GW energy and the total energy reservoir of the bubble (corresponding to the latent heat in the true vacuum region inside the bubble) scales as $ t$. Back-reaction effects (which we did not analyze) should become important if the bubble continues to live and expand for a time $t>t_{\rm br}$, with $t_{\rm br}\approx (16\pi^5)m_{\rm pl}^2 R_0^3$.

Assuming that percolation occurs before back-reaction becomes important, the produced GW energy spectrum peaks at  frequency $f_{*\rm peak}\approx1/R_0$. If the phase transition takes place in a radiation-dominated epoch with temperature $T_*$, with $g_*$ relativistic degrees of freedom in the plasma, the peak frequency observable today is $f_{\rm peak}\approx550~{\rm GHz}/\left(g_*^{\frac{1}{3}}R_0T_*\right)$. This is much higher than current and planned GW detector sensitivities, meaning that the spectrum would appear blue: either $\propto f$ or $\propto f^3$, depending if the experiment can observe above or below a break frequency corresponding to the onset of quadrupole domination. Unfortunately, however, simple estimates suggest that the signal falls below the sensitivity of even the most ambitious currently planned experiments, like the BBO experiment. 

Our analysis leaves room to many additional questions, for instance: are there models in which the back-reaction time is shorter than the duration of the phase transition? what happens in this case? what is the prediction of the thermal fluctuation analogue of our quantum fluctuation process? what happens if plasma friction slows down the bubble wall, in comparison to our scenario in which the bubble wall traversed the usual Rindler orbit? Finally, we have only been able to calculate GW production from the late stage of the evolution of the bubble, after it expands significantly. It would be interesting to tackle the problem of bubble fluctuation at nucleation. This problem would require going beyond the semiclassical approximation.

\acknowledgments
We thank Gui Pimentel, Wolfram Ratzinger, Antonio Riotto, Omri Rosner, Giovanni Villadoro, and Matias Zaldarriaga for useful discussions. KB is supported by ISF grant 1784/20 and by MINERVA grant 714123.

\begin{appendix}
\section{Minkowski parametrization}\label{app:chi}

Our perturbative treatment of wall fluctuations is seemingly unreliable after $t_{\rm NL}$, defined in \eqref{eq:tNL}, because of the exponential superhorizon growth of $\zeta$. However, as we saw in section \ref{sec:nl}, this is an artifact of trying to describe a fixed displacement in $r$ direction using the hyperbolic coordinates. Below, we will show that the above results are valid beyond $t_{\rm NL}$ by using the Minkowski parametrization of the wall. We define $\chi$ via
\be
r_w(t,\theta,\vphi) = \sqrt{R_0^2+t^2} + \chi(t,\theta,\vphi).
\ee
Substituting in the full action
\be
S = \sigma\int dt d\Omega \left[-r_w^2 \sqrt{1+ (\d r_w)^2} +\frac{r_w^3}{R_0}\right],
\ee
and expanding to quadratic order in $\chi$ results in
\be
S^{(2)} =\frac{1}{2}\sigma \int dt d\Omega \sqrt{1+\frac{t^2}{R_0^2}} \left[\frac{(R_0^2+t^2)^2}{R_0^2} \dot\chi^2 -|\nabla_a\chi|^2+2 \chi^2\right].
\ee
This can be quantized in a similar way as $\zeta$. First, expand in spherical basis
\be
\chi(t,\theta,\vphi) = \sum_{l,m} \chi_{lm}(t) Y_{lm}(\theta,\vphi),
\ee
and introduce ladder operators for each coefficient:
\be
\chi_{lm}(t) = a_{lm} f_l(t) + a_{lm}^\dagger f_l^*(t) ,
\ee
where $f_l(t)$ is a solution to 
\be
\d_t[(R_0^2+t^2)^{5/2}\dot f_l ] + R_0^2\sqrt{R_0^2+t^2}[l(l+1)-2] f_l = 0.
\ee
Which solution is decided by analytic continuation to imaginary time and requiring regularity at the tip of the resulting 3-sphere, at $t = -i R_0$. The result is
\be
f_l(t) =-\frac{\cos(l\pi)\Gamma\left(-l-\frac{1}{2}\right)\sqrt{\Gamma(l-1)\Gamma(3+l)} R_0^4}{2^{l+1}\pi \sqrt{\sigma R_0} t^4} \left(1+\frac{R_0^2}{t^2}\right)^{(l-1)/2}
  {_2F_1}\left(\frac{3+l}{2},\frac{4+l}{2},\frac{3}{2}+l,1+\frac{R_0^2}{(t-i \epsilon)^2}\right).
\ee
The normalization was fixed by taking the late time limit
\be\label{lateChi}\begin{split}
  f_l(t\gg l) =\sqrt{\frac{4 \Gamma(l-1)}{\pi \sigma R_0\Gamma(l+3)}} &\left( 1+\frac{(l(l+1) - 2)R_0^2}{4t^2}+\cdots +i \frac{\pi \Gamma(l+3)R_0^4}{32\Gamma(l-1) t^4}+\cdots\right),\end{split}
\ee
and imposing canonical commutation relation between $\chi$ and its conjugate momentum. We can check that expanding 
\be
r_w = \sqrt{R_0^2 e^{2\zeta} + t^2},
\ee
to linear order in $\zeta$ and substituting \eqref{eq:flate} reproduces the first two terms in \eqref{lateChi}.

The unit normal to the wall at first order in $\chi$ is
\be
d\xi= \left (-\frac{t}{R_0} -\left(1+\frac{t^2}{R_0^2}\right)^{3/2}\dot\chi,\hat x^i \sqrt{1+\frac{t^2}{R_0^2}}\left(1+ \frac{t \sqrt{R_0^2+t^2}}{R_0^2}\dot \chi\right)-\d_i\theta^a \nabla_a \chi\right).
\ee
This agrees with the expression in terms of $\zeta$
\be
d\xi = \left( -\frac{t}{R_0} (1-\zeta)-\sqrt{1+\frac{t^2}{R_0^2}}\d_\tau\zeta,
\hat x^i \left( \sqrt{1+\frac{t^2}{R_0^2}(1-2\zeta)} +\frac{t}{R_0}\d_\tau \zeta\right)- R_0 \d_i\theta^a \d_a \zeta\right),
\ee
after using \eqref{eq:flate} and \eqref{lateChi}. This implies that in the late time limit we will get the same $T_{\mu\nu}$ as in section \ref{sec:GW} and the rest of analysis remains identical. However, the description in terms of $\chi$ remains perturbative at large $t$.   

\section{Microscopic derivation}\label{a:mic}

Here we give an alternative derivation of the GW source term from bubble fluctuations, computing the scalar field EMT directly from the microscopic definition.

The thin-wall bubble is characterized by a scalar field profile of the form~\cite{Coleman:1977py,Callan:1977pt,Coleman:1980aw}
\be\phi_b(x)&\approx&a\,\tanh\left[\frac{\mu}{2}\left(\sqrt{r^2-t^2}-R\right)\right],\ee
where $r=|{\bf x}|$. This configuration provides an approximate solution to the classical equation of motion in the limit $\mu R\gg1$. In this limit, the field profile is real up to exponentially small complex oscillations in the timelike region inside the bubble, suppressed by\footnote{It is curious that these imaginary parts of the field are only suppressed by the constant factor $e^{-R\mu}$, namely, the field in the interior of the bubble does not become ``more and more classical" as the bubble expands in real time. One wonders if there are other interesting quantum effects that become relevant away from the thin-wall limit.} $\sim e^{-R\mu}$.   
It is natural to consider hyperbolic coordinates 
\be r&=&\rho\cosh\tau,\;\;\;\;{\bf x}=r\hat n,\;\;\;t=\rho\sinh\tau,\ee
that, for real $\rho$ and $\tau$, cover the part of Minkowski spacetime contained in the spacelike region w.r.t. the center of the bubble $x^\mu=(0,\vec 0)$. 
With this, $\phi_b=a\tanh\left[\frac{\mu}{2}\left(\rho-R\right)\right]$. 
We take the Lorentzian metric to be flat (up to, later on, GWs) 
\be ds^2&=&-g_{\mu\nu}dx^\mu dx^\nu=-dt^2+dr^2+r^2d\Omega^2=d\rho^2+\rho^2\left(-d\tau^2+\cosh^2\tau d\Omega^2\right),\ee
where $d\Omega^2=\sin\theta^2d\varphi^2+d\theta^2$. 
The Lorentzian action is
\be\label{eq:S} S&=&\int d^4x\sqrt{-g}\left[\frac{1}{2}g^{\mu\nu}\frac{\partial\phi}{\partial x^\mu}\frac{\partial\phi}{\partial x^\nu}-V\right]\\
&\sim&\int d\tau\cosh^2\tau\int d\Omega\,\int d\rho\rho^3\left[\frac{1}{2\rho^2}\left(\frac{\partial\phi}{\partial\tau}\right)^2-\frac{1}{2}\left(\frac{\partial\phi}{\partial\rho}\right)^2-\frac{1}{2\rho^2\cosh^2\tau\sin^2\theta}\left(\frac{\partial\phi}{\partial\varphi}\right)^2-\frac{1}{2\rho^2\cosh^2\tau}\left(\frac{\partial\phi}{\partial\theta}\right)^2-V\right].\no
\ee
We use $\sim$ in the second line because it integrates the Lagrangian density only over the spacelike region w.r.t. the center of the bubble. In the thin-wall limit, the omitted timelike region is interior to the bubble wall, so it contributes a constant. Notably, when we get to the effective action for fluctuations on the bubble wall, we will only need to capture the wall world-volume, for which the action above is adequate.

The potential responsible for the thin wall solution is~\cite{Coleman:1977py,Callan:1977pt,Coleman:1980aw}
\be V&=&\frac{\mu^2}{8a^2}\left(\phi^2-a^2\right)^2+\frac{\epsilon}{2a}\left(\phi-a\right).\ee
The unperturbed bubble satisfies $\frac{1}{2}\left(\frac{\partial\phi_b}{\partial\rho}\right)^2=\frac{\mu^2}{8a^2}\left(\phi_b^2-a^2\right)^2$, and integration over $\rho$ gives $\int d\rho\rho^3\left[-\frac{1}{2}\left(\frac{\partial\phi_b}{\partial\rho}\right)^2-\frac{\mu^2}{8a^2}\left(\phi_b^2-a^2\right)^2\right]=-\int d\rho\rho^3\frac{\mu^2}{4a^2}\left(\phi_b^2-a^2\right)^2=-\frac{2a^2\mu R^3}{3}\equiv-\sigma R^3$. 
The $\epsilon$ piece gives $-\int d\rho\rho^3\frac{\epsilon}{2a}\left(\phi_b-a\right)=\frac{\epsilon R^4}{4}\left(1+\mathcal{O}\left((R\mu)^{-2}\right)\right)$. The unperturbed action is therefore
\be S_b&=&\int d\tau\cosh^2\tau\int d\Omega\left(-\sigma R^3+\frac{\epsilon}{4}R^4\right),\ee
extremized by 
\be\label{R0se} R_0&=&\frac{3\sigma}{\epsilon}.\ee

We would have obtained the same result had we done the calculation using an Euclidean action, defined via the analytic continuation $\tau\to-i\mathcal{T}$, with $\mathcal{T}$ real. This would change $r\to\rho\cos\mathcal{T}$ and $t\to-i\rho\sin\mathcal{T}$, and swap the sign of $\left(\partial\phi/\partial\tau\right)^2$, but since the unperturbed bubble profile only depends on $\rho$, and $\partial\phi_b/\partial\tau=0$, the action would retain the same form and yield the same result when extremized w.r.t. R.

For $\mu R\gg1$, low-energy excitations of the wall propagate only in the transverse directions $y^\alpha=(\tau,\theta,\varphi)$. These excitations can be described by a 3D Nambu-Goto EFT. We parameterize these excitations by letting 
\be R&=&R_0e^\zeta,
\ee
where $\zeta=\zeta(y)$. To $\mathcal{O}(\zeta^2)$, this gives
\be S^{(2)}&=&\int d\tau\cosh^2\tau\int d\Omega\,\left[\int d\rho\rho\frac{R^2}{2}\left(\frac{\partial\phi_b}{\partial\rho}\right)^2\left(\left(\frac{\partial\zeta}{\partial \tau}\right)^2-\frac{1}{\cosh^2\tau\sin^2\theta}\left(\frac{\partial\zeta}{\partial\varphi}\right)^2-\frac{1}{\cosh^2\tau}\left(\frac{\partial\zeta}{\partial\theta}\right)^2\right)-\sigma R^3+\frac{\epsilon R^4}{4}\right]\no\\
%
%
&=&\frac{\sigma R_0^3}{2}\int d\tau\cosh^2\tau\int d\Omega\,\left[\left(\frac{\partial\zeta}{\partial \tau}\right)^2-\frac{1}{\cosh^2\tau\sin^2\theta}\left(\frac{\partial\zeta}{\partial\varphi}\right)^2-\frac{1}{\cosh^2\tau}\left(\frac{\partial\zeta}{\partial\theta}\right)^2+3\zeta^2\right]+\mathcal{O}\left(\zeta^3\right)+{\rm Const}.\no\ee
$\zeta$ has a tachyonic mass.

\subsection{EMT}
The EMT is
\be T_{\mu\nu}(x)&=&\partial_\mu\phi\partial_\nu\phi+\eta_{\mu\nu}\left(V(\phi)-\frac{1}{2}\left(\partial_\alpha\phi\right)^2\right).
\ee
The fluctuations $\zeta(\tau,\varphi,\theta)$ correspond to field fluctuations
\be\phi(x)&\approx&{a}\tanh\left[\frac{\mu}{2}\left(\sqrt{r^2-t^2}-R_0\right)\right]-\frac{a\mu R_0\zeta}{2\cosh^2\left[\frac{\mu}{2}\left(\sqrt{r^2-t^2}-R_0\right)\right]}+\mathcal{O}\left(\zeta^2\right).\ee
The spatial gradient of the field is
\be\partial_i\phi(x)&\approx&\frac{r}{\sqrt{r^2-t^2}}\frac{a\mu \,\hat n_i}{2\cosh^2\left[\frac{\mu}{2}\left(\sqrt{r^2-t^2}-R_0\right)\right]}
-\frac{a\mu R_0\,\partial_i\zeta}{2\cosh^2\left[\frac{\mu}{2}\left(\sqrt{r^2-t^2}-R_0\right)\right]}
\no\\&+&\frac{a\mu^2\,\tanh\left[\frac{\mu}{2}\left(\sqrt{r^2-t^2}-R_0\right)\right]}{2\cosh^2\left[\frac{\mu}{2}\left(\sqrt{r^2-t^2}-R_0\right)\right]}\frac{R_0r\,\hat n_i\,\zeta}{\sqrt{r^2-t^2}}+\mathcal{O}\left(\zeta^2\right).
\ee
The term $\propto\eta_{\mu\nu}$ will drop in the projection onto the transverse-traceless GW source, so we focus on $\Pi_{ij}=\partial_i\phi\partial_j\phi$. 
We will need the EMT to first order in $\zeta$, so we truncate $\partial_i\phi$ at the same order. 
We find
\be\Pi_{ij}&=&\sigma\frac{r^2}{r^2-t^2}\frac{3\mu}{8\cosh^4\left[\frac{\mu}{2}\left(\sqrt{r^2-t^2}-R_0\right)\right]}\hat n_i\hat n_j\\
&-&\sigma\frac{r}{\sqrt{r^2-t^2}}\frac{3\mu}{8\cosh^4\left[\frac{\mu}{2}\left(\sqrt{r^2-t^2}-R_0\right)\right]}R_0\left(\hat n_i\partial_j\zeta+\hat n_j\partial_i\zeta\right)
\no\\&+&\sigma\frac{r}{\sqrt{r^2-t^2}}\left(\frac{r}{\sqrt{r^2-t^2}}\frac{3\mu^2\,\tanh\left[\frac{\mu}{2}\left(\sqrt{r^2-t^2}-R_0\right)\right]}{4\cosh^4\left[\frac{\mu}{2}\left(\sqrt{r^2-t^2}-R_0\right)\right]}\right)R_0\zeta\,\hat n_i\hat n_j
+\mathcal{O}\left(\zeta^2\right).\no\ee
To clarify the choice of presentation, note the following representation of the Dirac delta function,
\be\lim_{\mu\to\infty}\frac{r}{R_0}\frac{3\mu}{8\cosh^4\left[\frac{\mu}{2}\left(\sqrt{r^2-t^2}-R_0\right)\right]}&=&\delta\left(r-\sqrt{R_0^2+t^2}\right).\ee
With this in mind we define
\be\delta_w\left(r-\sqrt{R_0^2+t^2}\right)&=&\frac{r}{R_0}\frac{3\mu}{8\cosh^4\left[\frac{\mu}{2}\left(\sqrt{r^2-t^2}-R_0\right)\right]}.\ee
For the purpose of spatial integration at constant $t$, we will treat $\delta_w$ as a Dirac delta. We also note the identity
\be\frac{r}{\sqrt{r^2-t^2}}\frac{3\mu^2\,\tanh\left[\frac{\mu}{2}\left(\sqrt{r^2-t^2}-R_0\right)\right]}{4\cosh^4\left[\frac{\mu}{2}\left(\sqrt{r^2-t^2}-R_0\right)\right]}&=&-\partial_r\left[\frac{R_0}{r}\delta_w\left(r-\sqrt{R_0^2+t^2}\right)\right],\ee
and remind the identification of surface tension
\be\sigma&=&\frac{2a^2\mu}{3}.\ee
With these definitions we can write
\be\Pi_{ij}&=&\sigma\delta_w\left(r-\sqrt{R_0^2+t^2}\right)\frac{R_0\,r}{r^2-t^2}\hat n_i\hat n_j
-\sigma\delta_w\left(r-\sqrt{R_0^2+t^2}\right)\frac{R^2_0}{\sqrt{r^2-t^2}}\left(\hat n_i\partial_j\zeta+\hat n_j\partial_i\zeta\right)
\no\\&-&\sigma R_0\zeta\,\hat n_i\hat n_j\,\frac{r}{\sqrt{r^2-t^2}}\partial_r\left[\frac{R_0}{r}\delta_w\left(r-\sqrt{R_0^2+t^2}\right)\right]+\mathcal{O}\left(\zeta^2\right)
.\ee

The $(+)$ helicity traceless-transverse projector for GWs with ${\bf k}||\hat z$ is
\be\varepsilon_+^{ij}&=&\frac{1}{2}\left(\hat x+i\hat y\right)^i\left(\hat x+i\hat y\right)^j.\ee
In polar coordinates, the relevant projections are
\be\epsilon^{ij}_+\hat n_i\hat n_j&=&\frac{\sin^2\theta\,e^{2i\varphi}}{2},\\
\epsilon^{ij}_+\hat n_i\partial_j&=&\frac{\sin\theta \,e^{2i\varphi}}{2r}\left(\cos\theta\partial_\theta+\frac{i}{\sin\theta}\partial_\varphi+\sin\theta \,r\partial_r\right).
\ee

With this, the GW source term is 
\be\label{eq:Tpm2app}T_+\left(t,{\bf k}\right)&=&\int d^3xe^{i{\bf kx}}\varepsilon_+^{ij}\Pi_{ij}\left(t,{\bf x}\right)\\
&=&\frac{\sigma}{2}\int drr^2\,\frac{R_0\,r}{r^2-t^2}\delta_w\left(r-\sqrt{R_0^2+t^2}\right)\int d\theta\sin^3\theta\,\int d\varphi\,e^{ikr\cos\theta+2i\varphi}\no\\
&-&\sigma R_0\int drr\,\frac{R_0}{\sqrt{r^2-t^2}}\delta_w\left(r-\sqrt{R_0^2+t^2}\right)\int d\theta\sin^2\theta\,\int d\varphi\,e^{ikr\cos\theta+2i\varphi}\left(\cos\theta\partial_\theta+\frac{i}{\sin\theta}\partial_\varphi+\sin\theta \,r\partial_r\right)\zeta\no\\
&+&\frac{\sigma R_0}{2}\int dr\,\frac{R_0}{r}\delta_w\left(r-\sqrt{R_0^2+t^2}\right)\int d\theta\sin^3\theta\,\int d\varphi\,\partial_r\left[\frac{r^3}{\sqrt{r^2-t^2}}\,e^{ikr\cos\theta+2i\varphi}\zeta\right]\no\\
&=&-\frac{\sigma R_0r_w}{2}\int d\theta\sin^3\theta\,\int d\varphi\,e^{ikr_w\cos\theta+2i\varphi}\left(-\frac{r_wt}{R_0^2}\d_\tau\zeta
+\left(4+\frac{t^2}{R_0^2}+ikr_w\cos\theta\right)\zeta
\right).\no\ee

This reproduces Eq.~(\ref{eq:Tpm2}).

\section{Asymptotics}\label{app:l}
Below, we will first discuss the limits $kt\gg l^2$ and $kt\ll l$, and then the limit $kt\sim l\gg 1$, which determines the large $t$ behavior of $E_{\rm GW}$. 

\subsection{$x = kt\gg l^2$}
In this limit, one of the terms in the expression \eqref{eq:Tp} for $T_{+,l}$ dominates,
\be
I_l(x)\approx 2i x^3 \int_{-1}^1 dc\ c (1-c^2)\,e^{ic x}P_{l,2}(c).\no
\ee
When $x\gg l^2$, the integral is dominated by $x\sim \pm 1$, where we can use the asymptotic behavior of $P_{l,2}$ to get
\be
I_l(x) \approx -4 i^l \frac{\Gamma(l+3)}{\Gamma(l-1)}\cos\left(x-\frac{l \pi}{2}\right).
\ee
Substituting in \eqref{eq:hpmA} gives
\be\label{eq:Hl2}
H_l(x) \approx -2 i^l \frac{\Gamma(l+3)}{\Gamma(l-1)} x\sin\left(x-\frac{l\pi}{2}\right).
\ee
Therefore, the UV part of \eqref{eq:El} is
\be
\expect{E_{{\rm GW},l}}_{\rm UV} \approx \frac{G\sigma t^3}{R_0^3} \int^\infty d\ln x \frac{2l}{\pi^3x}.
\ee
\subsection{$x=kt\ll l$}
In this limit, we can expand $e^{ixc}$ in the integrand of $I_l(x)$. There is a nonzero overlap between $(1-c^2)P_{l,2}(c)$ and $c^{n}$ if $l-n =0$ mod $2$, and $n\geq l-2$. Therefore, at small $x$, $I_l(x) \propto x^{l}$. We can use 
\be
\int_{-1}^1 dc\ c^n P_l(c) = \frac{2^l \left((-1)^{l+n}+1\right) \Gamma (n+1) \Gamma \left(\frac{l+n}{2}+1\right)}{\Gamma (l+n+2) \Gamma \left(\frac{n-l}{2}+1\right)},
\ee
and that $P_{l,2}(c) = (1-c^2)P_l''(c)$ to find its coefficient:
\be
\int_{-1}^1 dc\ c^{l-2} (1-c^2) P_{l,2}(c) = \frac{\sqrt{\pi } 2^{-l-1} \Gamma (l-1)}{\Gamma \left(l+\frac{3}{2}\right)}.
\ee
Substituting in \eqref{eq:hpmA} and Taylor expanding $\sin(x-w)$ gives
\be
H_l(x) \propto x^{l+2}.
\ee
It is the $|H'_l(x)|^2\propto x^{2l+2}$ term in \eqref{eq:El} that dominates $E_l$ in the IR:
\be\label{eq:Euv}
\expect{E_{{\rm GW},l}}_{\rm IR} \approx \frac{G\sigma t^3}{R_0^3} \int_0 d\ln x\ c_{\rm IR} x^{2l-1}.
\ee
Because of the steep growth, the IR contribution is subdominant in the large $l$ limit. So it is not useful to calculate $c_{\rm IR}$. 
\subsection{$x=kt \sim l\gg 1$}
For any $l$, there is an exact, but unwieldy, expression for $I_l(x)$ in terms of Bessel functions. At large $l$, it simplifies to 
\be
I_l(x)\approx f(x) j_{l+4}(x) + g(x) j'_{l+4}(x)
\ee
where
\be 
f(x) = -i^l\frac{2 l^6 \left(8 l^4-8 l^2 x^2+x^4\right)}{x^4},\qquad
g(x) = -i^l\frac{2 l^6 \left(8 l^3 x-4 l x^3\right)}{x^4},
\ee
and $j_l(x)$ is the spherical Bessel function. It has a classical turning point at $x_t = \sqrt{l(l+1)}$. At large $l$, it decays rapidly in the forbidden region, i.e. within $\Delta x \sim l^{1/3}$ when $x<x_t$. When $x-x_t\gg l^{1/3}$, it rapidly oscillates and is well approximated via WKB:
\be
j_l(x) \approx \frac{\sin(x+\Phi(x))}{x\sqrt{\omega(x)}}
\ee
where
\be
\omega(x) \equiv \sqrt{1-\frac{l(l+1)}{x^2}},\qquad \Phi(x) \equiv -\int_x^\infty dy\ [\omega_l(y)-1]-\frac{l\pi}{2}.
\ee
Since $E_l$ is dominated by the contribution within a region of size $\Delta x \sim l\gg l^{1/3}$, we can calculate its asymptotic behavior by using the above WKB expression. In this regime, derivatives of sine and cosine are much larger than the prefactors, hence
\be
I_l(x) \approx \frac{1}{x}
\left[\frac{f(x)}{\sqrt{\omega(x)}} \sin(x+\Phi(x))+g(x) \sqrt{\omega(x)} \cos(x+\Phi(x))\right].
\ee
Since the prefactors are slowly varying, such that $\frac{d}{d x}\sim \frac{1}{x}$, we can approximate
\be
\int^x dw\ \frac{f(w)}{w\sqrt{\omega(w)}} \sin(x-w) \sin(w+\Phi(w))
\approx  \frac{f(x)}{2x\sqrt{\omega(x)}}\sin(x+\Phi(x)) 
\left(\frac{1}{2+\Phi'(x)}-\frac{1}{\Phi'(x)}\right),
\ee
with a similar expression for the cosine integral, resulting in
\be\label{eq:Hl}
H_l(x) \approx \frac{x}{l^2} \left[\frac{f(x)}{\sqrt{\omega(x)}} \sin(x+\Phi(x)) - g(x) \sqrt{\omega(x)} \cos(x+\Phi(x))\right].
\ee
Note that when $x\gg l$ the above expression matches the one in \eqref{eq:Hl2}, even though the latter was derived assuming $x\gg l^2$. Hence, the power spectrum in \eqref{eq:Euv} holds for all $k\gg l/t$. 

The UV spectrum $P_l(s)\propto 1/s$ implies that $E_{{\rm GW},l}$ is dominated by $x\sim l$ and we have to keep all terms in \eqref{eq:Hl}. So to leading order in $l$,
\be
\left\langle E_{{\rm GW},l}(t)\right\rangle &\approx & \frac{ G \sigma t^3 l}{\pi^3R_0^3} \int_{l}^\infty 
\frac{dx}{x^4}\frac{(2 x^2 - l^2)}{\sqrt{1-\frac{l^2}{x^2}}}
= \frac{3 G \sigma t^3}{4 \pi^2R_0^3}.
\ee

\end{appendix}

\bibliography{ref}
\bibliographystyle{utphys}

\end{document}